\documentclass[aps,pra, twocolumn]{revtex4-1}
\usepackage{graphicx,color,amsmath,amssymb,enumerate,verbatim,cancel, relsize }
\usepackage{amsmath}
\usepackage{multirow}
\pdfoutput=1

\newcommand{\bra}[1]{\langle #1|}
\newcommand{\ket}[1]{|#1\rangle}
\newcommand{\inner}[2]{\langle #1 | #2 \rangle}

\begin{document}

\title{Tripartite non-classical correlations without backwards in time signalling }
\date{\today}
\author{Tom Purves} \affiliation{H.H. Wills Physics Laboratory, University of Bristol, Tyndall Avenue, Bristol, BS8 1TL, U.K.}
\author{Anthony J. Short} \affiliation{H.H. Wills Physics Laboratory, University of Bristol, Tyndall Avenue, Bristol, BS8 1TL, U.K.}

\begin{abstract}
We investigate the relationship between no backwards in time signaling and classically causal correlations. We discover that unlike the case for two parties, the no backwards in time signaling paradigm for three parties is not enough to ensure that correlations can be reproduced with a classical causal ordering. We demonstrate this with an explicit example. We also generalise some existing results for linear two-time states to multi-party scenarios.  
\end{abstract}

\maketitle

\section{Introduction} 

Recently, there has been growing interest in indefinite causal order in quantum theory \cite{Aharonov1990, Oreshkov2012, Chiribella2013, Leifer2013, Baumeler2014, Branciard2016, Allen2017}, for which no classical ordering of events can explain the results. This may arise for many reasons, for example due to quantum switches (in which the order of two operations is controlled by a quantum bit) \cite{Chiribella2013, qswitch, witnessing},  superpositions of causal order arising from a  quantum theory of gravity \cite{hardyqgrav, Hardy2009, Zych2017}, or post-selection \cite{Aharonov1964, Aharonov1991, Aharonov2009,  Silva2014, connectingprocesandtwotime, polytope}. 

An interesting situation to consider is one in which a number of parties carry out experiments within individual laboratories, each of which obeys standard quantum theory (with a fixed causal order), whilst the connections between laboratories are more exotic. The most general object representing such connections is a \emph{process matrix} \cite{Oreshkov2012, Branciard2016, Oreshkov2016, Abbott2016, Baumeler2016, computationwithcausality}. Process matrices can lead to correlations between laboratories which defy any classically causal explanation, in an analogous way to that in which Bell-inequality-violating correlations defy local explanation \cite{bell}.

In a recent paper \cite{polytope}, a particular class of experiments was considered, in which each party performs a fixed measurement followed by a chosen transformation. In this scenario,  no-backwards-in-time-signalling (NBTS) correlations are considered, for which each party individually sees results consistent with the normal flow of time within their laboratory (i.e. their measurement results do not depend on which transformation they later performed). This leads to NBTS-conditions which are analogous to the spatial no-signalling conditions for Bell-type scenarios. It was  shown for two parties that a connection between the laboratories involving post-selection satisfies the NBTS-conditions if and only if it can be represented by a process matrix (or equivalently a linear two-time state). In this paper, we extend these results to multiple parties. 

Surprisingly, the correlations that could be obtained in the two-party NBTS scenario were limited to the classically causal set, despite the existence of consistent non-classical possibilities. This raises the interesting question of whether process matrices always lead to classical correlations in the NBTS-scenario. If this were the case it could lead to interesting insights about the nature of quantum causality. In this paper, however, we show that this is not the case, by giving an explicit counterexample for three parties.

\section{The NBTS paradigm}
\begin{figure}
\centering 
\includegraphics[width=0.29\textwidth]{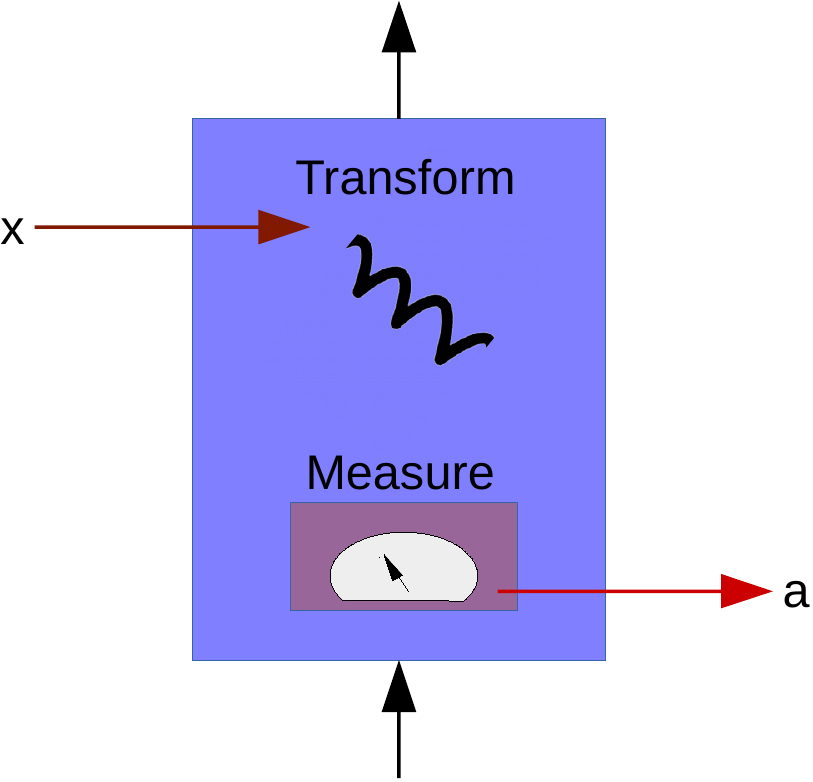}
\caption{A laboratory which obeys the NBTS paradigm. Experimenters in this lab always make measurements before transformations.\label{fig:NBTS-scenario}}
\end{figure}
We first clarify the NBTS scenario. Any protocol that fits within this framework contains a number of labeled laboratories, with each laboratory obeying a well defined procedure. The procedure (illustrated in figure \ref{fig:NBTS-scenario}) is always of the form:

\begin{enumerate}
\item A system enters laboratory. 
\item A fixed measurement of the received system is made, and the classical result is recorded as an output. 
\item A classical input is received. This can be thought of  either as randomly generated, or provided by an independent external agent. 
\item A transformation of the system is made, which may depend on the received input.
\item The system exists the lab. 
\end{enumerate}
If time in the labs is flowing normally we expect that outputs of the laboratories cannot depend on the inputs. Considering  a single party, the  conditional probability distribution obeyed by her input $x$ and output $a$  therefore satisfies
\begin{equation}
p(a|x)=p(a)
\end{equation}
which we call the NBTS condition \footnote{For a single party, this is equivalent to $a$ and $x$ being uncorrelated}. 

Now consider three parties, Alice, Bob and Charlie, with each party obeying the NBTS procedure. Alice gets input $x$, and outputs $a$. Bob and Charlie get inputs $y$ and $z$, and output $b$ and $c$ respectively. We expect the probability distribution for Alice's output, obtained by marginalizing $p(a,b,c|x,y,z)$ over Bob and Charlie's outputs, to also be independent of her input, because the NBTS scenario imposes that the generation of output happens strictly before her input. More explicitly, we expect
\begin{equation}
p_A(a|x,y,z) \equiv \sum_{b,c}  p(a,b,c|x,y,z)
\end{equation}
to be independent of $x$. Hence;
\begin{equation} \label{eq:NBTS1} 
p_A(a|x,y,z)=p_A(a|x',y,z) \equiv p_A(a|y,z)
\end{equation}
for all $a,x, x',y,z$. We also need to same to hold for Bob, as he should also conclude that there is no backwards in time signaling. Thus, Bobs marginal distribution should follow
\begin{equation}
p_B(b|x,y,z)=p_B(b|x,z), 
\end{equation}
and similarly for Charlie,
\begin{equation}\label{eq:NBTS2} 
p_C(c|x,y,z)=p_C(c|x,y).
\end{equation}

We may describe the permitted probability distributions as a convex polytope in probability-space \cite{Abbott2016, polytope}. Here we consider the case where all inputs and outputs are binary for simplicity. For three parties, the situation is described by the 64 co-ordinates $p(a,b,c|x,y,z)$ for all possible values of the inputs and outputs. We demand that our coordinates are probabilities, and as such each coordinate varies between $0$ and $1$, and respect normalization,
\begin{align}
0\leq p(a,b,c|x,y,z) \leq 1 \qquad
\sum_{a, b, c} p(a,b,c|x,y,z)=1 .
\end{align}
We also require that the probability distribution obeys the NBTS conditions \eqref{eq:NBTS1}-\eqref{eq:NBTS2} on each party. This gives a convex set which we refer to as the NBTS-polytope (which plays an analogous role to the non-signalling polytpe in non-locality).  We  also define a classical polytope that sits within the NBTS polytope that contains probabilities that could be achieved classically (analogous to the local polytope), in which  systems can be represented by classical random variables, and the laboratories are arranged either with a deterministic classical ordering, or a mixture of such orderings. Deterministic classical strategies will sit at vertices of such a polytope, and hence it can be generated in vertex representation by considering all  such classical strategies. For example, suppose  Alice goes first. This means that her output can only ever be a constant bit, as it occurs before any input. There is however some freedom to choose the subsequent ordering based on her input. For example, suppose that  Alice's output system encodes her input bit, and that when $x=0$ the system is next given to Bob and then Charlie, whereas when $x=1$ the system is given to Charlie then Bob. Such a strategy is described by the formula
\begin{align}
p(a,b,c|0,y,z)=\begin{cases}
1 \text{ if } &a= \alpha, b=\beta, c= \delta y \oplus \gamma\\
0 & \text { otherwise}
\end{cases}
\\
p(a,b,c|1,y,z)=\begin{cases}
1 \text{ if } &a= \alpha, c=\lambda, b=\nu z \oplus \mu\\
0 & \text { otherwise.}
\end{cases}
\end{align}
where $\alpha, \beta, \gamma, \delta, \lambda,\mu, \nu$ are bits and $\oplus$ denotes addition modulo 2. We then construct the full polytope by considering all strategies of this form, and constructing the convex hull. 

Finally, we also define a quantum polytope, which again is analogous to the polytope of the same name in the non-locality scenario. Strategies within the quantum polytope are allowed to use process matrices outside the laboratories and standard quantum theory inside the laboratories. Hence; the quantum polytope sits in the middle of the size hierarchy. In \cite{polytope} it was shown that for the case of two parties the quantum polytope and the classical polytope are identical in the NBTS scenario.  While there are plenty of exotic ways to wire together two parties, we find that any resultant physical correlations are no different to those generated by Alice going first or Bob going first, or a mixture of the two. We may then ask, is the NBTS paradigm enough to ensure such a classical description? 

In section \ref{sec:protocol} we will show that this is not the case, by giving an explicit quantum protocol for three parties that generates an extreme point of the NBTS polytope, that is strictly not contained within the classical polytope. 

\section{Linear Two-Time States}

In \cite{connectingprocesandtwotime} it was shown that any two-party process matrix can be simulated within quantum theory, given the power of post-selection. In particular, process matrices correspond to a particular set of pre- and post-selected states called linear two-time states.

 The formalism we use to describe pre- and post-selected states is developed from \cite{Aharonov1964, Aharonov2009, Silva2014} and described in detail in \cite{connectingprocesandtwotime}. Key elements of the formalism are that all states, measurement operators and channels are represented as vectors in Hilbert spaces and their dual spaces, while time evolution is represented in a symmetric way by the connections between these vectors.
 Different vectors can be combined via the $\bullet$ operation, which connects vectors within a Hilbert space 
and its dual to give a scalar (i.e. $_{\mathcal{A}}\bra{\psi} \bullet \ket{\phi}^{\mathcal{A}} = \inner{\psi}{\phi}$), whilst combining vectors
in different Hilbert spaces as a tensor product. As in relativity, we perform contractions between vectors with the same label raised and lowered. We  call a vector $\eta$ (with labels $\eta^{A_i B_i C_i}_{A_o B_o C_o} $) a  linear two-time state  for three parties if it obeys the relation
  \begin{equation}
 p(a,b,c|x,y,z)=\left( J(a|x)^{A_o}_{A_i}  \otimes K(b|y)^{B_o}_{B_i} \otimes L(c|z)^{C_o}_{C_i}  \right) \bullet  \eta \label{eqn:defining}
 \end{equation}
where $J(a|x)^{A_o}_{A_i}$ represents a standard quantum measurement channel\footnote{i.e. a trace non-increasing completely positive map.} between Alice's input $A_i$ and output $A_o$  which depends on $x$ and has output $a$, and similarly for the other two parties. We note \eqref{eqn:defining} is a linear function of the state and measurements. 

In  appendix \ref{ap:lineartwotime} we show explicitly that the linear two-time state and process matrix formalisms are equivalent for any number of parties; given a process matrix there will always be a linear two-time state that is isomorphic. In appendix \ref{ap:linearNBTS} we also show that imposing the NBTS conditions on a pre- and post-selected quantum state for all choices of the parties' measurements and transformations enforces that the state can be represented by a linear two-time state. These results were shown in \cite{connectingprocesandtwotime, polytope} for two parties, but  we generalize them to any number of parties.
\begin{figure*}[ht]
\centering
\includegraphics[width=\textwidth]{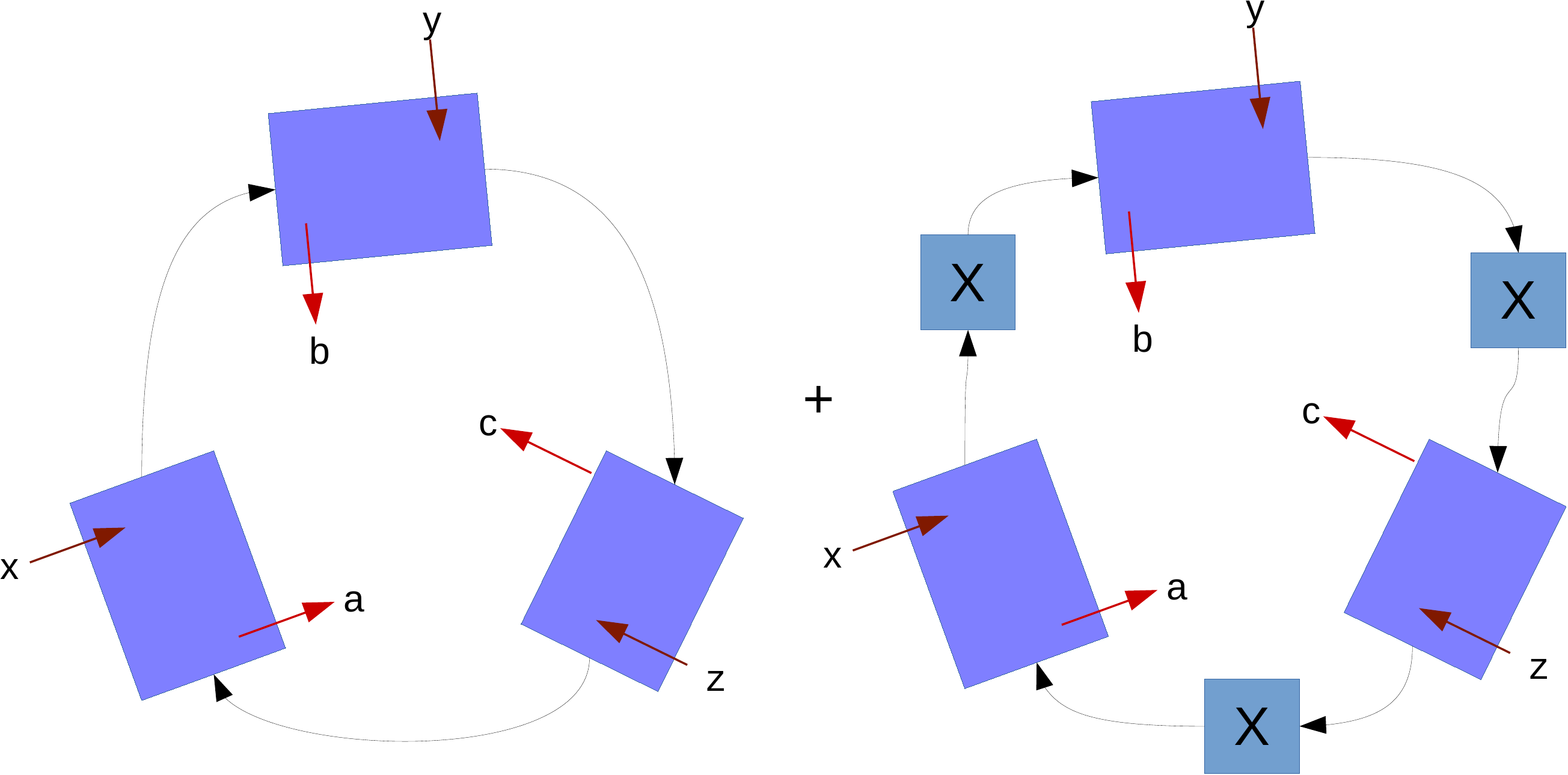}
\caption{Pictorial representation of the setup that $\eta$ models. The black lines connecting laboratories represent channels with a $z$-basis measurement. The gates labeled $X$ are flip gates. Notice that in the frame of each laboratory the output is produced before the input is received.\label{fig:eta}}
\end{figure*}
\section{The Protocol} \label{sec:protocol} 
In this section, we construct a tripartite NBTS-scenario which cannot be explained classically. We  use the same wiring of the laboratories as that given by Baumeler et al. in \cite{Baumeler2014} to show that classical process matrices can violate causal inequalities, but a new protocol that respects the NBTS-scenario  and only requires a binary input for each party. Alice, Bob, and Charlie are sat in their closed labs. Outside of the labs, they have no idea of how they are connected together. They each perform the following protocol:  At some point a qubit enters their lab. They first measure the qubit  in the computational basis, then  receive a classical input bit. If this input bit is 0, they pass the qubit out of the door of their labs, and if it is a 1, they flip the qubit (by applying the gate $X=\ket{0}\bra{1} + \ket{1}\bra{0}$) and then pass it out. The action of Alice, whose input and output spaces are labelled by $A_i$ and $A_o$, and who received input $x$ and obtained output $a$ in her measurement, can be represented by the vectors 
\begin{align}
M(a|x)_{A_i}^{A_o} &= |a\oplus x \rangle^{\mathcal{A}_o} \!\otimes\!\,_{\mathcal{A}_i} \!\langle a |  \otimes |a \rangle^{{\mathcal{A}_i}^\dagger} \!\otimes \!\,_{\mathcal{A}_o^\dagger}\!\langle a \oplus x|.
\end{align}
We denote primitive Hilbert spaces with calligraphic letters, and full Hilbert spaces with Roman labeling, i.e $H^{A}=H^{\mathcal{A}}\otimes H_{\mathcal{A}^\dagger}$. The laboratories have been wired together in rather odd way. The labs are connected in a cyclic fashion, as shown in figure \ref{fig:eta}, with probability $1/2$ that there is an external $z$ basis measurement between them all. There is also $1/2$ probability that they are all connected with a $z$ basis measurement followed by an $X$ gate. This can be represented by the two-time state 
\begin{equation}
\eta = \frac{1}{2}(M_{A_o} ^{B_i}\otimes M_{B_o} ^{C_i} \otimes M_{C_o} ^{A_i} + \bar{M}_{A_o} ^{B_i} \otimes \bar{M}_{B_o} ^{C_i} \otimes \bar{M}_{C_o} ^{A_i})
\end{equation}
where the operators $M$ and $\bar{M}$ are given as
\begin{align}
M_A^B=M(0|0)_A^B + M(1|0)_A^B \\
\bar{M}_A^B=M(0|1)_A^B + M(1|1)_A^B.
\end{align}
One may think at first glance that this setup does not respect the NBTS paradigm; it seems like there is a direct path between Alice's output and her input. However, a careful consideration of the scenario shows that  a maximally random state will always enter each of the parties laboratories, and as such the NBTS conditions are respected. Furthermore, we show in appendix  \ref{append:etaistwotime} that $\eta$ is a linear two-time state, and hence will satisfy the NBTS conditions regardless of the specific measurements and transformations used by Alice, Bob and Charlie. 

In order to calculate the measurement probabilities for this scenario, we use \eqref{eqn:defining} to get
\begin{equation}
p(a,b,c|x,y,z)=(M(a|x)_{A_i}^{A_o} \otimes  M(b|y)_{B_i}^{B_o} \otimes  M(c|z)_{C_i}^{C_o})\bullet  \eta.
\end{equation}
Evaluating this expression we arrive at the compact probability relation
\begin{align}
p(a,b,c|x,y,z)=\frac{1}{2}\left(\delta_{b,a\oplus x}\delta_{c,b\oplus y}\delta_{a,c\oplus z}+\delta_{b,a\oplus \bar{x}}\delta_{c,b\oplus \bar{y}}\delta_{a,c\oplus \bar{z}}\right)
\end{align}
where we use a bar above a bit to represent its negation (e.g. $\bar{x}=x \oplus 1$). This generates the probabilities in table \ref{tab:prob}. 
\begin{table}[h]
\centering
\begin{tabular}{|l|l|l|l|l|l|l|l|l|l|}
\hline
\multicolumn{2}{|l|}{\multirow{2}{*}{$p(a,b,c|x,y,z)$}} & \multicolumn{8}{c|}{$a,b,c$}                    \\ \cline{3-10} 
\multicolumn{2}{|l|}{}                                & 000 & 001 & 010 & 100 & 011 & 101 & 110 & 111 \\ \hline
\multirow{8}{*}{$x,y,z$}              & 000             & 1/2 & 0   & 0   & 0   & 0   & 0   & 0   & 1/2 \\ \cline{2-10} 
                                    & 001             & 0   & 0   & 1/2 & 0   & 0   & 1/2 & 0   & 0   \\ \cline{2-10} 
                                    & 010             & 0   & 0   & 0   & 1/2 & 1/2 & 0   & 0   & 0   \\ \cline{2-10} 
                                    & 100             & 0   & 1/2 & 0   & 0   & 0   & 0   & 1/2 & 0   \\ \cline{2-10} 
                                    & 011             & 0   & 1/2 & 0   & 0   & 0   & 0   & 1/2 & 0   \\ \cline{2-10} 
                                    & 101             & 0   & 0   & 0   & 1/2 & 1/2 & 0   & 0   & 0   \\ \cline{2-10} 
                                    & 110             & 0   & 0   & 1/2 & 0   & 0   & 1/2 & 0   & 0   \\ \cline{2-10} 
                                    & 111             & 1/2 & 0   & 0   & 0   & 0   & 0   & 0   & 1/2 \\ \hline
\end{tabular}
\caption{Table of probabilities generated by the protocol.}
\label{tab:prob}
\end{table}

We note that it carries some interesting relations, namely we have that:
\begin{align}
p(a,b,c|x,y,z)=&p(\bar{a},\bar{b},\bar{c}|x,y,z)=p(a,b,c|\bar{x},\bar{y},\bar{z})\\
=&p(b,c,a|y,z,x)=p(a,\bar{b},c|x,y,\bar{z}).
\end{align}
We can use these relations to show that this probability table cannot be generated by any classical strategy, or affine combination of such strategies. Let $p_i$ be some deterministic classical strategy that could contribute to the affine combination that makes up $p$, i.e. $p=\sum_i \lambda_i p_i$, with $0\leq \lambda \leq 1$. Then, consider a particular choice of $a,b,c,x,y,z$ such that 
\begin{align}
p_i(a,b,c|x,y,z)>0,  
\end{align}
Given that $ p(a,b,c|x,y,z) \geq p_i(a,b,c|x,y,z)$, this is only possible if  $p(a,b,c|x,y,z)=0.5$ (the only non-zero value in the probability table), which also implies 
\begin{align} 
p(a,\bar{b},c|x,y,\bar{z})=0.5 \text{ and } p(\bar{a},b,\bar{c}|x,y,\bar{z})=0.5.
\end{align}
Hence, using normalization
\begin{equation}
p(a,b,c|x,y,\bar{z})=0. \label{eq:equalzero1} 
\end{equation}
The cyclic symmetry may be used to derive similar relations for the logical not of the other inputs, so that
\begin{align}
p(a,b,c|x,\bar{y},z)=0 \quad \text{   and } \quad p(a,b,c|\bar{x},y,z)=0. \label{eq:equalzero2} 
\end{align}
Yet, the deterministic classical strategy $p_i$ has some party that goes last when the inputs are $x,y,z$. The probability distribution must be the same when the last party's input is flipped (as it occurs after all outputs). Hence $p_i(a,b,c|x,y,z)>0$ implies
\begin{align}
p_i(a,b,c|\bar{x},y,z)+p_i(a,b,c|x,\bar{y},z)+p_i(a,b,c|x,y,\bar{z})>0,
\end{align}
as one of the three terms must be positive. This further implies 
\begin{align}
p(a,b,c|\bar{x},y,z)+p(a,b,c|x,\bar{y},z)+p(a,b,c|x,y,\bar{z})>0,
\end{align}
which creates a contradiction  with \eqref{eq:equalzero1} and \eqref{eq:equalzero2}. Hence we cannot write $p$ as a mixture of classical strategies.

 Any distribution in the polytope must satisfy the NBTS equalities and normalization. The remaining constraints are all positivity inequalities, which are saturated when $p(a,b,c|x,y,z)=0$. The probability distribution given in Table \ref{tab:prob} then saturates a total of 64 linearly independent equalities in a 64 dimensional space, hence it must be the unique point satisfying all of those equalities. The only way it would be possible to mix two points together to obtain this distribution would be for them both to violate the NBTS or normalisation equalities, or for one of them to violate a saturated positivity condition. However both of these methods utilize points outside the polytope. It follows that the distribution produced by the protocol must  
 be an extreme point of the NBTS set (and also an extreme point of the quantum polytope). 

\section{Discussion and Conclusions}

We have given an explicit protocol that cannot be realized with any classical causal order, yet can be implemented with a linear two time state (or process matrix), or within a more general theory that allows for indefinite causal orderings. Moreover, we have shown that this is  an extreme point of the NBTS polytope. This is in stark contrast to the two party NBTS scenario, where it has been shown that no such non-classical correlations exist. It is interesting to note that  the state, measurements and transformations used in the protocol only involve $z$-basis measurements and $X$-gates, and could therefore be obtained using classical stochastic processes and post-selection. An interesting question is whether all quantum correlations for the NBTS scenario have such stochastic analogues.  

In this paper, we have focussed on the case of indefinite relative timings between the three parties. In \cite{polytope}, cases in which the timings of two parties are definite are also considered, and it would be interesting to investigate these cases further for three or more parties. 

Another interesting question is what, if any, additional constraints or scenarios can be considered beyond the NBTS case which are sufficient to ensure a classical causal explanation for any pre- and post-selected quantum procedure? 
 
 \acknowledgements
 
The authors acknowlage helpful conversations with P. Skrzypczyk. TP acknowledges support from the EPSRC. 

\bibliography{references.bib} 

\begin{thebibliography}{25}%
\makeatletter
\providecommand \@ifxundefined [1]{%
 \@ifx{#1\undefined}
}%
\providecommand \@ifnum [1]{%
 \ifnum #1\expandafter \@firstoftwo
 \else \expandafter \@secondoftwo
 \fi
}%
\providecommand \@ifx [1]{%
 \ifx #1\expandafter \@firstoftwo
 \else \expandafter \@secondoftwo
 \fi
}%
\providecommand \natexlab [1]{#1}%
\providecommand \enquote  [1]{``#1''}%
\providecommand \bibnamefont  [1]{#1}%
\providecommand \bibfnamefont [1]{#1}%
\providecommand \citenamefont [1]{#1}%
\providecommand \href@noop [0]{\@secondoftwo}%
\providecommand \href [0]{\begingroup \@sanitize@url \@href}%
\providecommand \@href[1]{\@@startlink{#1}\@@href}%
\providecommand \@@href[1]{\endgroup#1\@@endlink}%
\providecommand \@sanitize@url [0]{\catcode `\\12\catcode `\$12\catcode
  `\&12\catcode `\#12\catcode `\^12\catcode `\_12\catcode `\%12\relax}%
\providecommand \@@startlink[1]{}%
\providecommand \@@endlink[0]{}%
\providecommand \url  [0]{\begingroup\@sanitize@url \@url }%
\providecommand \@url [1]{\endgroup\@href {#1}{\urlprefix }}%
\providecommand \urlprefix  [0]{URL }%
\providecommand \Eprint [0]{\href }%
\providecommand \doibase [0]{http://dx.doi.org/}%
\providecommand \selectlanguage [0]{\@gobble}%
\providecommand \bibinfo  [0]{\@secondoftwo}%
\providecommand \bibfield  [0]{\@secondoftwo}%
\providecommand \translation [1]{[#1]}%
\providecommand \BibitemOpen [0]{}%
\providecommand \bibitemStop [0]{}%
\providecommand \bibitemNoStop [0]{.\EOS\space}%
\providecommand \EOS [0]{\spacefactor3000\relax}%
\providecommand \BibitemShut  [1]{\csname bibitem#1\endcsname}%
\let\auto@bib@innerbib\@empty
\bibitem [{\citenamefont {{Aharonov}}\ \emph {et~al.}(1990)\citenamefont
  {{Aharonov}}, \citenamefont {{Anandan}}, \citenamefont {{Popescu}},\ and\
  \citenamefont {{Vaidman}}}]{Aharonov1990}%
  \BibitemOpen
  \bibfield  {author} {\bibinfo {author} {\bibfnamefont {Y.}~\bibnamefont
  {{Aharonov}}}, \bibinfo {author} {\bibfnamefont {J.}~\bibnamefont
  {{Anandan}}}, \bibinfo {author} {\bibfnamefont {S.}~\bibnamefont
  {{Popescu}}}, \ and\ \bibinfo {author} {\bibfnamefont {L.}~\bibnamefont
  {{Vaidman}}},\ }\href {\doibase 10.1103/PhysRevLett.64.2965} {\bibfield
  {journal} {\bibinfo  {journal} {\prl}\ }\textbf {\bibinfo {volume} {64}},\
  \bibinfo {pages} {2965} (\bibinfo {year} {1990})}\BibitemShut {NoStop}%
\bibitem [{\citenamefont {{Oreshkov}}\ \emph {et~al.}(2012)\citenamefont
  {{Oreshkov}}, \citenamefont {{Costa}},\ and\ \citenamefont
  {{Brukner}}}]{Oreshkov2012}%
  \BibitemOpen
  \bibfield  {author} {\bibinfo {author} {\bibfnamefont {O.}~\bibnamefont
  {{Oreshkov}}}, \bibinfo {author} {\bibfnamefont {F.}~\bibnamefont {{Costa}}},
  \ and\ \bibinfo {author} {\bibfnamefont {{\v{C}}.}~\bibnamefont
  {{Brukner}}},\ }\href {\doibase 10.1038/ncomms2076} {\bibfield  {journal}
  {\bibinfo  {journal} {Nature Communications}\ }\textbf {\bibinfo {volume}
  {3}},\ \bibinfo {eid} {1092} (\bibinfo {year} {2012})},\ \Eprint
  {http://arxiv.org/abs/1105.4464} {arXiv:1105.4464 [quant-ph]} \BibitemShut
  {NoStop}%
\bibitem [{\citenamefont {Chiribella}\ \emph {et~al.}(2013)\citenamefont
  {Chiribella}, \citenamefont {D'Ariano}, \citenamefont {Perinotti},\ and\
  \citenamefont {Valiron}}]{Chiribella2013}%
  \BibitemOpen
  \bibfield  {author} {\bibinfo {author} {\bibfnamefont {G.}~\bibnamefont
  {Chiribella}}, \bibinfo {author} {\bibfnamefont {G.~M.}\ \bibnamefont
  {D'Ariano}}, \bibinfo {author} {\bibfnamefont {P.}~\bibnamefont {Perinotti}},
  \ and\ \bibinfo {author} {\bibfnamefont {B.}~\bibnamefont {Valiron}},\ }\href
  {\doibase 10.1103/PhysRevA.88.022318} {\bibfield  {journal} {\bibinfo
  {journal} {Phys. Rev. A}\ }\textbf {\bibinfo {volume} {88}},\ \bibinfo
  {pages} {022318} (\bibinfo {year} {2013})}\BibitemShut {NoStop}%
\bibitem [{\citenamefont {Leifer}\ and\ \citenamefont
  {Spekkens}(2013)}]{Leifer2013}%
  \BibitemOpen
  \bibfield  {author} {\bibinfo {author} {\bibfnamefont {M.~S.}\ \bibnamefont
  {Leifer}}\ and\ \bibinfo {author} {\bibfnamefont {R.~W.}\ \bibnamefont
  {Spekkens}},\ }\href {\doibase 10.1103/PhysRevA.88.052130} {\bibfield
  {journal} {\bibinfo  {journal} {Phys. Rev. A}\ }\textbf {\bibinfo {volume}
  {88}},\ \bibinfo {pages} {052130} (\bibinfo {year} {2013})}\BibitemShut
  {NoStop}%
\bibitem [{\citenamefont {Baumeler}\ and\ \citenamefont
  {Wolf}(2014)}]{Baumeler2014}%
  \BibitemOpen
  \bibfield  {author} {\bibinfo {author} {\bibfnamefont {{\"A}.}~\bibnamefont
  {Baumeler}}\ and\ \bibinfo {author} {\bibfnamefont {S.}~\bibnamefont
  {Wolf}},\ }in\ \href {\doibase 10.1109/ISIT.2014.6874888} {\emph {\bibinfo
  {booktitle} {2014 IEEE International Symposium on Information Theory}}}\
  (\bibinfo {year} {2014})\ pp.\ \bibinfo {pages} {526--530}\BibitemShut
  {NoStop}%
\bibitem [{\citenamefont {Branciard}\ \emph {et~al.}(2016)\citenamefont
  {Branciard}, \citenamefont {Araújo}, \citenamefont {Feix}, \citenamefont
  {Costa},\ and\ \citenamefont {\v{C}aslav Brukner}}]{Branciard2016}%
  \BibitemOpen
  \bibfield  {author} {\bibinfo {author} {\bibfnamefont {C.}~\bibnamefont
  {Branciard}}, \bibinfo {author} {\bibfnamefont {M.}~\bibnamefont {Araújo}},
  \bibinfo {author} {\bibfnamefont {A.}~\bibnamefont {Feix}}, \bibinfo {author}
  {\bibfnamefont {F.}~\bibnamefont {Costa}}, \ and\ \bibinfo {author}
  {\bibnamefont {\v{C}aslav Brukner}},\ }\href
  {http://stacks.iop.org/1367-2630/18/i=1/a=013008} {\bibfield  {journal}
  {\bibinfo  {journal} {New Journal of Physics}\ }\textbf {\bibinfo {volume}
  {18}},\ \bibinfo {pages} {013008} (\bibinfo {year} {2016})}\BibitemShut
  {NoStop}%
\bibitem [{\citenamefont {{Allen}}\ \emph {et~al.}(2017)\citenamefont
  {{Allen}}, \citenamefont {{Barrett}}, \citenamefont {{Horsman}},
  \citenamefont {{Lee}},\ and\ \citenamefont {{Spekkens}}}]{Allen2017}%
  \BibitemOpen
  \bibfield  {author} {\bibinfo {author} {\bibfnamefont {J.-M.~A.}\
  \bibnamefont {{Allen}}}, \bibinfo {author} {\bibfnamefont {J.}~\bibnamefont
  {{Barrett}}}, \bibinfo {author} {\bibfnamefont {D.~C.}\ \bibnamefont
  {{Horsman}}}, \bibinfo {author} {\bibfnamefont {C.~M.}\ \bibnamefont
  {{Lee}}}, \ and\ \bibinfo {author} {\bibfnamefont {R.~W.}\ \bibnamefont
  {{Spekkens}}},\ }\href {\doibase 10.1103/PhysRevX.7.031021} {\bibfield
  {journal} {\bibinfo  {journal} {Physical Review X}\ }\textbf {\bibinfo
  {volume} {7}},\ \bibinfo {eid} {031021} (\bibinfo {year} {2017})}\BibitemShut
  {NoStop}%
\bibitem [{\citenamefont {{Ebler}}\ \emph {et~al.}(2018)\citenamefont
  {{Ebler}}, \citenamefont {{Salek}},\ and\ \citenamefont
  {{Chiribella}}}]{qswitch}%
  \BibitemOpen
  \bibfield  {author} {\bibinfo {author} {\bibfnamefont {D.}~\bibnamefont
  {{Ebler}}}, \bibinfo {author} {\bibfnamefont {S.}~\bibnamefont {{Salek}}}, \
  and\ \bibinfo {author} {\bibfnamefont {G.}~\bibnamefont {{Chiribella}}},\
  }\href {\doibase 10.1103/PhysRevLett.120.120502} {\bibfield  {journal}
  {\bibinfo  {journal} {Physical Review Letters}\ }\textbf {\bibinfo {volume}
  {120}},\ \bibinfo {eid} {120502} (\bibinfo {year} {2018})},\ \Eprint
  {http://arxiv.org/abs/1711.10165} {arXiv:1711.10165 [quant-ph]} \BibitemShut
  {NoStop}%
\bibitem [{\citenamefont {{Ara{\'u}jo}}\ \emph {et~al.}(2015)\citenamefont
  {{Ara{\'u}jo}}, \citenamefont {{Branciard}}, \citenamefont {{Costa}},
  \citenamefont {{Feix}}, \citenamefont {{Giarmatzi}},\ and\ \citenamefont
  {{Brukner}}}]{witnessing}%
  \BibitemOpen
  \bibfield  {author} {\bibinfo {author} {\bibfnamefont {M.}~\bibnamefont
  {{Ara{\'u}jo}}}, \bibinfo {author} {\bibfnamefont {C.}~\bibnamefont
  {{Branciard}}}, \bibinfo {author} {\bibfnamefont {F.}~\bibnamefont
  {{Costa}}}, \bibinfo {author} {\bibfnamefont {A.}~\bibnamefont {{Feix}}},
  \bibinfo {author} {\bibfnamefont {C.}~\bibnamefont {{Giarmatzi}}}, \ and\
  \bibinfo {author} {\bibfnamefont {{\v C}.}~\bibnamefont {{Brukner}}},\ }\href
  {\doibase 10.1088/1367-2630/17/10/102001} {\bibfield  {journal} {\bibinfo
  {journal} {New Journal of Physics}\ }\textbf {\bibinfo {volume} {17}},\
  \bibinfo {eid} {102001} (\bibinfo {year} {2015})},\ \Eprint
  {http://arxiv.org/abs/1506.03776} {arXiv:1506.03776 [quant-ph]} \BibitemShut
  {NoStop}%
\bibitem [{\citenamefont {{Hardy}}(2007)}]{hardyqgrav}%
  \BibitemOpen
  \bibfield  {author} {\bibinfo {author} {\bibfnamefont {L.}~\bibnamefont
  {{Hardy}}},\ }\href {\doibase 10.1088/1751-8113/40/12/S12} {\bibfield
  {journal} {\bibinfo  {journal} {Journal of Physics A Mathematical General}\
  }\textbf {\bibinfo {volume} {40}},\ \bibinfo {pages} {3081} (\bibinfo {year}
  {2007})},\ \Eprint {http://arxiv.org/abs/gr-qc/0608043} {gr-qc/0608043}
  \BibitemShut {NoStop}%
\bibitem [{\citenamefont {Hardy}(2009)}]{Hardy2009}%
  \BibitemOpen
  \bibfield  {author} {\bibinfo {author} {\bibfnamefont {L.}~\bibnamefont
  {Hardy}},\ }\enquote {\bibinfo {title} {Quantum gravity computers: On the
  theory of computation with indefinite causal structure},}\ in\ \href
  {\doibase 10.1007/978-1-4020-9107-0_21} {\emph {\bibinfo {booktitle} {Quantum
  Reality, Relativistic Causality, and Closing the Epistemic Circle: Essays in
  Honour of Abner Shimony}}}\ (\bibinfo  {publisher} {Springer Netherlands},\
  \bibinfo {address} {Dordrecht},\ \bibinfo {year} {2009})\ pp.\ \bibinfo
  {pages} {379--401}\BibitemShut {NoStop}%
\bibitem [{\citenamefont {{Zych}}\ \emph {et~al.}(2017)\citenamefont {{Zych}},
  \citenamefont {{Costa}}, \citenamefont {{Pikovski}},\ and\ \citenamefont
  {{Brukner}}}]{Zych2017}%
  \BibitemOpen
  \bibfield  {author} {\bibinfo {author} {\bibfnamefont {M.}~\bibnamefont
  {{Zych}}}, \bibinfo {author} {\bibfnamefont {F.}~\bibnamefont {{Costa}}},
  \bibinfo {author} {\bibfnamefont {I.}~\bibnamefont {{Pikovski}}}, \ and\
  \bibinfo {author} {\bibfnamefont {C.}~\bibnamefont {{Brukner}}},\ }\href@noop
  {} {\bibfield  {journal} {\bibinfo  {journal} {ArXiv e-prints}\ } (\bibinfo
  {year} {2017})},\ \Eprint {http://arxiv.org/abs/1708.00248} {arXiv:1708.00248
  [quant-ph]} \BibitemShut {NoStop}%
\bibitem [{\citenamefont {Aharonov}\ \emph {et~al.}(1964)\citenamefont
  {Aharonov}, \citenamefont {Bergmann},\ and\ \citenamefont
  {Lebowitz}}]{Aharonov1964}%
  \BibitemOpen
  \bibfield  {author} {\bibinfo {author} {\bibfnamefont {Y.}~\bibnamefont
  {Aharonov}}, \bibinfo {author} {\bibfnamefont {P.~G.}\ \bibnamefont
  {Bergmann}}, \ and\ \bibinfo {author} {\bibfnamefont {J.~L.}\ \bibnamefont
  {Lebowitz}},\ }\href {\doibase 10.1103/PhysRev.134.B1410} {\bibfield
  {journal} {\bibinfo  {journal} {Phys. Rev.}\ }\textbf {\bibinfo {volume}
  {134}},\ \bibinfo {pages} {B1410} (\bibinfo {year} {1964})}\BibitemShut
  {NoStop}%
\bibitem [{\citenamefont {{Aharonov}}\ and\ \citenamefont
  {{Vaidman}}(1991)}]{Aharonov1991}%
  \BibitemOpen
  \bibfield  {author} {\bibinfo {author} {\bibfnamefont {Y.}~\bibnamefont
  {{Aharonov}}}\ and\ \bibinfo {author} {\bibfnamefont {L.}~\bibnamefont
  {{Vaidman}}},\ }\href {\doibase 10.1088/0305-4470/24/10/018} {\bibfield
  {journal} {\bibinfo  {journal} {Journal of Physics A Mathematical General}\
  }\textbf {\bibinfo {volume} {24}},\ \bibinfo {pages} {2315} (\bibinfo {year}
  {1991})}\BibitemShut {NoStop}%
\bibitem [{\citenamefont {Aharonov}\ \emph {et~al.}(2009)\citenamefont
  {Aharonov}, \citenamefont {Popescu}, \citenamefont {Tollaksen},\ and\
  \citenamefont {Vaidman}}]{Aharonov2009}%
  \BibitemOpen
  \bibfield  {author} {\bibinfo {author} {\bibfnamefont {Y.}~\bibnamefont
  {Aharonov}}, \bibinfo {author} {\bibfnamefont {S.}~\bibnamefont {Popescu}},
  \bibinfo {author} {\bibfnamefont {J.}~\bibnamefont {Tollaksen}}, \ and\
  \bibinfo {author} {\bibfnamefont {L.}~\bibnamefont {Vaidman}},\ }\href
  {\doibase 10.1103/PhysRevA.79.052110} {\bibfield  {journal} {\bibinfo
  {journal} {Phys. Rev. A}\ }\textbf {\bibinfo {volume} {79}},\ \bibinfo
  {pages} {052110} (\bibinfo {year} {2009})}\BibitemShut {NoStop}%
\bibitem [{\citenamefont {{Silva}}\ \emph {et~al.}(2014)\citenamefont
  {{Silva}}, \citenamefont {{Guryanova}}, \citenamefont {{Brunner}},
  \citenamefont {{Linden}}, \citenamefont {{Short}},\ and\ \citenamefont
  {{Popescu}}}]{Silva2014}%
  \BibitemOpen
  \bibfield  {author} {\bibinfo {author} {\bibfnamefont {R.}~\bibnamefont
  {{Silva}}}, \bibinfo {author} {\bibfnamefont {Y.}~\bibnamefont
  {{Guryanova}}}, \bibinfo {author} {\bibfnamefont {N.}~\bibnamefont
  {{Brunner}}}, \bibinfo {author} {\bibfnamefont {N.}~\bibnamefont {{Linden}}},
  \bibinfo {author} {\bibfnamefont {A.~J.}\ \bibnamefont {{Short}}}, \ and\
  \bibinfo {author} {\bibfnamefont {S.}~\bibnamefont {{Popescu}}},\ }\href
  {\doibase 10.1103/PhysRevA.89.012121} {\bibfield  {journal} {\bibinfo
  {journal} {Physical Review A}\ }\textbf {\bibinfo {volume} {89}},\ \bibinfo
  {eid} {012121} (\bibinfo {year} {2014})},\ \Eprint
  {http://arxiv.org/abs/1308.2089} {arXiv:1308.2089 [quant-ph]} \BibitemShut
  {NoStop}%
\bibitem [{\citenamefont {Silva}\ \emph {et~al.}(2017)\citenamefont {Silva},
  \citenamefont {Guryanova}, \citenamefont {Short}, \citenamefont {Skrzypczyk},
  \citenamefont {Brunner},\ and\ \citenamefont
  {Popescu}}]{connectingprocesandtwotime}%
  \BibitemOpen
  \bibfield  {author} {\bibinfo {author} {\bibfnamefont {R.}~\bibnamefont
  {Silva}}, \bibinfo {author} {\bibfnamefont {Y.}~\bibnamefont {Guryanova}},
  \bibinfo {author} {\bibfnamefont {A.~J.}\ \bibnamefont {Short}}, \bibinfo
  {author} {\bibfnamefont {P.}~\bibnamefont {Skrzypczyk}}, \bibinfo {author}
  {\bibfnamefont {N.}~\bibnamefont {Brunner}}, \ and\ \bibinfo {author}
  {\bibfnamefont {S.}~\bibnamefont {Popescu}},\ }\href
  {http://stacks.iop.org/1367-2630/19/i=10/a=103022} {\bibfield  {journal}
  {\bibinfo  {journal} {New Journal of Physics}\ }\textbf {\bibinfo {volume}
  {19}},\ \bibinfo {pages} {103022} (\bibinfo {year} {2017})}\BibitemShut
  {NoStop}%
\bibitem [{\citenamefont {{Guryanova}}\ \emph {et~al.}(2017)\citenamefont
  {{Guryanova}}, \citenamefont {{Silva}}, \citenamefont {{Short}},
  \citenamefont {{Skrzypczyk}}, \citenamefont {{Brunner}},\ and\ \citenamefont
  {{Popescu}}}]{polytope}%
  \BibitemOpen
  \bibfield  {author} {\bibinfo {author} {\bibfnamefont {Y.}~\bibnamefont
  {{Guryanova}}}, \bibinfo {author} {\bibfnamefont {R.}~\bibnamefont
  {{Silva}}}, \bibinfo {author} {\bibfnamefont {A.~J.}\ \bibnamefont
  {{Short}}}, \bibinfo {author} {\bibfnamefont {P.}~\bibnamefont
  {{Skrzypczyk}}}, \bibinfo {author} {\bibfnamefont {N.}~\bibnamefont
  {{Brunner}}}, \ and\ \bibinfo {author} {\bibfnamefont {S.}~\bibnamefont
  {{Popescu}}},\ }\href@noop {} {\bibfield  {journal} {\bibinfo  {journal}
  {ArXiv e-prints}\ } (\bibinfo {year} {2017})},\ \Eprint
  {http://arxiv.org/abs/1708.00669} {arXiv:1708.00669 [quant-ph]} \BibitemShut
  {NoStop}%
\bibitem [{\citenamefont {{Oreshkov}}\ and\ \citenamefont
  {{Giarmatzi}}(2016)}]{Oreshkov2016}%
  \BibitemOpen
  \bibfield  {author} {\bibinfo {author} {\bibfnamefont {O.}~\bibnamefont
  {{Oreshkov}}}\ and\ \bibinfo {author} {\bibfnamefont {C.}~\bibnamefont
  {{Giarmatzi}}},\ }\href {\doibase 10.1088/1367-2630/18/9/093020} {\bibfield
  {journal} {\bibinfo  {journal} {New Journal of Physics}\ }\textbf {\bibinfo
  {volume} {18}},\ \bibinfo {eid} {093020} (\bibinfo {year}
  {2016})}\BibitemShut {NoStop}%
\bibitem [{\citenamefont {{Abbott}}\ \emph {et~al.}(2016)\citenamefont
  {{Abbott}}, \citenamefont {{Giarmatzi}}, \citenamefont {{Costa}},\ and\
  \citenamefont {{Branciard}}}]{Abbott2016}%
  \BibitemOpen
  \bibfield  {author} {\bibinfo {author} {\bibfnamefont {A.~A.}\ \bibnamefont
  {{Abbott}}}, \bibinfo {author} {\bibfnamefont {C.}~\bibnamefont
  {{Giarmatzi}}}, \bibinfo {author} {\bibfnamefont {F.}~\bibnamefont
  {{Costa}}}, \ and\ \bibinfo {author} {\bibfnamefont {C.}~\bibnamefont
  {{Branciard}}},\ }\href {\doibase 10.1103/PhysRevA.94.032131} {\bibfield
  {journal} {\bibinfo  {journal} {Physical Review A}\ }\textbf {\bibinfo
  {volume} {94}},\ \bibinfo {eid} {032131} (\bibinfo {year}
  {2016})}\BibitemShut {NoStop}%
\bibitem [{\citenamefont {{Baumeler}}\ and\ \citenamefont
  {{Wolf}}(2016)}]{Baumeler2016}%
  \BibitemOpen
  \bibfield  {author} {\bibinfo {author} {\bibfnamefont {{\"A}.}~\bibnamefont
  {{Baumeler}}}\ and\ \bibinfo {author} {\bibfnamefont {S.}~\bibnamefont
  {{Wolf}}},\ }\href {\doibase 10.1088/1367-2630/18/3/035014} {\bibfield
  {journal} {\bibinfo  {journal} {New Journal of Physics}\ }\textbf {\bibinfo
  {volume} {18}},\ \bibinfo {eid} {035014} (\bibinfo {year}
  {2016})}\BibitemShut {NoStop}%
\bibitem [{\citenamefont {{Ara{\'u}jo}}\ \emph {et~al.}(2017)\citenamefont
  {{Ara{\'u}jo}}, \citenamefont {{Gu{\'e}rin}},\ and\ \citenamefont
  {{Baumeler}}}]{computationwithcausality}%
  \BibitemOpen
  \bibfield  {author} {\bibinfo {author} {\bibfnamefont {M.}~\bibnamefont
  {{Ara{\'u}jo}}}, \bibinfo {author} {\bibfnamefont {P.~A.}\ \bibnamefont
  {{Gu{\'e}rin}}}, \ and\ \bibinfo {author} {\bibfnamefont
  {{\"A}.}~\bibnamefont {{Baumeler}}},\ }\href {\doibase
  10.1103/PhysRevA.96.052315} {\bibfield  {journal} {\bibinfo  {journal}
  {\pra}\ }\textbf {\bibinfo {volume} {96}},\ \bibinfo {eid} {052315} (\bibinfo
  {year} {2017})},\ \Eprint {http://arxiv.org/abs/1706.09854} {arXiv:1706.09854
  [quant-ph]} \BibitemShut {NoStop}%
\bibitem [{\citenamefont {Bell}(1964)}]{bell}%
  \BibitemOpen
  \bibfield  {author} {\bibinfo {author} {\bibfnamefont {J.~S.}\ \bibnamefont
  {Bell}},\ }\href {\doibase 10.1103/PhysicsPhysiqueFizika.1.195} {\bibfield
  {journal} {\bibinfo  {journal} {Physics Physique Fizika}\ }\textbf {\bibinfo
  {volume} {1}},\ \bibinfo {pages} {195} (\bibinfo {year} {1964})}\BibitemShut
  {NoStop}%
\bibitem [{Note1()}]{Note1}%
  \BibitemOpen
  \bibinfo {note} {For a single party, this is equivalent to $a$ and $x$ being
  uncorrelated}\BibitemShut {NoStop}%
\bibitem [{Note2()}]{Note2}%
  \BibitemOpen
  \bibinfo {note} {I.e. a trace non-increasing completely positive
  map.}\BibitemShut {Stop}%
\end{thebibliography}%


%

\onecolumngrid
\appendix
\newpage

\section{Linear Two-Time States and Process Matrices give rise to the same probabilities} \label{ap:lineartwotime} 
We give a general proof that process matrices and linear two time states exist that give the same probabilities. Furthermore, following \cite{connectingprocesandtwotime} we observe a natural isomorphism between process matrices and the linear two-time states. we begin with the probability rule for process matrices from \cite{Oreshkov2016}.
\begin{align}
P(a_1,...,a_N)&=\text{tr}\bigg[ W^{A^1_1A^1_2...A^N_1 A^N_2}\bigg(\bigotimes_{i=1}^{N} M_{a_i}^{A^i_1 A^i_2}\bigg)\bigg] \nonumber \\
&=\text{tr}\bigg[\sum_{\mu_1...\mu_N} W^{A^1_1A^1_2...A^N_1 A^N_2} \bigg( \bigotimes_{i=1}^{N} \bigg[ (\mathbb{I}\otimes \hat{E}^{[i]}_{\mu_i, a_i})|\Phi^+ \rangle\langle \Phi^+|(\mathbb{I}\otimes \hat{E}^{[i] \dagger}_{\mu_i, a_i})\bigg]^T\bigg)\bigg] \nonumber\\
&=\text{tr}\bigg[\sum_{\mu_1...\mu_N} (W^{A^1_1A^1_2...A^N_1 A^N_2})^T \bigg( \bigotimes_{i=1}^{N} \bigg[ (\mathbb{I}\otimes \hat{E}^{[i]}_{\mu_i, a_i})|\Phi^+ \rangle\langle \Phi^+|(\mathbb{I}\otimes \hat{E}^{[i] \dagger}_{\mu_i, a_i})\bigg]\bigg)\bigg]\nonumber\\
&=\text{tr}\bigg[\sum_{k_1...k_{2N} \atop l_1...l_{2N}}\sum_{\mu_1...\mu_N} w_{k_1...k_{2N},l_1...l_{2N}}|l_1...l_{2N} \rangle\langle k_1...k_{2N}| \bigg( \bigotimes_{i=1}^{N} \bigg[ (\mathbb{I}\otimes \hat{E}^{[i]}_{\mu_i, a_i})|\Phi^+ \rangle\langle \Phi^+|(\mathbb{I}\otimes \hat{E}^{[i] 
\dagger}_{\mu_i, a_i})\bigg]\bigg)\bigg]\nonumber\\
&=\sum_{k_1...k_{2N} \atop l_1...l_{2N}}\sum_{\mu_1...\mu_N}w_{k_1...k_{2N},l_1...l_{2N}} \bigg( \prod_{i=1}^{N}\, \text{tr} \bigg[ |l_{2i-1}l_{2i}\rangle\langle k_{2i-1}k_{2i}|(\mathbb{I}\otimes \hat{E}^{[i]}_{\mu_i, a_i}) \sum_{t_i, u_i} |t_i t_i\rangle\langle u_i u_i| (\mathbb{I}\otimes \hat{E}^{[i] 
\dagger}_{\mu_i, a_i})\bigg]\bigg)\nonumber\\
&= \sum_{k_1...k_{2N}}\sum_{l_1...l_{2N}}\sum_{\mu_1...\mu_N}w_{k_1...k_{2N},l_1...l_{2N}} \bigg( \prod_{i=1}^{N}\,  \langle k_{2i}|\hat{E}^{[i]}_{\mu_i, a_i}|k_{2i-1}\rangle \langle l_{2i-1}|\hat{E}^{[i] \dagger}_{\mu_i, a_i}|l_{2i}\rangle \bigg)\nonumber\\
&= \bigg(\sum_{k_1...k_{2N}}\sum_{l_1...l_{2N}}w_{k_1...k_{2N},l_1...l_{2N}}    \,_{\mathcal{A}_2^1}\langle k_2| \otimes |k_1\rangle^{\mathcal{A}_1^1} \otimes _{\mathcal{A}_1^{1 \dagger}}\langle l_1| \otimes |l_2\rangle^{\mathcal{A}_2^{1 \dagger}} \otimes ...\nonumber\\ &... \otimes  _{\mathcal{A}_2^{N}}\langle k_{2N}| \otimes |k_{2N-1}\rangle^{\mathcal{A}_1^{N}} \otimes _{\mathcal{A}_1^{N \dagger}}\langle l_{2N-1}|\otimes  |l_{2N}\rangle ^{\mathcal{A}_2^{N \dagger}}\bigg) \bullet \bigg( \sum_{\mu_1} \hat{E}^{[1]}_{\mu_1, a_1}\otimes \hat{E}^{[1] \dagger}_{\mu_1,a_1} \otimes ... \otimes \sum_{\mu_N} \hat{E}^{[N]}_{\mu_N, a_N}\otimes \hat{E}^{[N] \dagger}_{\mu_N, a_N}\bigg)\nonumber\\
&=\eta_W  \bullet (J_{a_1}^{[1]}  \otimes...  \otimes J_{a_N}^{[N]})
\end{align}
which is equivalent to the probability rule for linear two-time states as quoted in the main text in \eqref{eqn:defining}. Furthermore, note that $\eta$ is `positive' (in the sense required to be a valid two-time state \cite{connectingprocesandtwotime}) if and only if $W$ is a positive operator. The isomorphism between $W$ and $\eta$ is obtained by expressing the former as a vector in the basis chosen for the transpose and flipping bras and kets on the output spaces. 

\section{Connection between linearity of two-time states and the NBTS conditions} \label{ap:linearNBTS} 

To see that a linear two-time state always satisfies the NBTS conditions, note that it is equivalent to a process matrix. These are defined such that the phenomena inside individual laboratories obeys standard quantum theory (without post-selection) and thus cannot lead to backwards in time signalling. Alternatively, one can construct a multiparty analogue of equation (30) in \cite{polytope} and use a similar argument to that in theorem 2 of \cite{connectingprocesandtwotime}. 

Our task then is to prove that satisfying the NBTS conditions implies that a pre- and post-selected quantum state can be represent by a linear two-time state, for which we use a multiparty generalisation of the argument in theorem 2 of \cite{connectingprocesandtwotime}. We first consider the marginal state of the first party, given by 
\begin{equation}
\eta_{A_1}=(J^{[2]} \otimes \ldots \otimes  J^{[N]}) \bullet  \eta 
\end{equation}
where $J^{[k]}$ is the channel for the $k$'th party summed over their output, which may depend on their input $x_k$, and could be an arbitrary channel. It was shown in \cite{connectingprocesandtwotime} that the only single party marginal states that satisfy NBTS take the form of a product with the identity operator on the output
\begin{equation}
\eta_{A_1} = (\rho^{A_1^1} \otimes \mathbb{I}_{A_2^1})
\end{equation}
corresponding to states with no post-selection, or trivial post selection. Hence,
\begin{align}
(J^{[1]} \otimes J^{[2]} \otimes \ldots \otimes  J^{[N]}) \bullet  \eta &= (J^{[1]})_{A_1^1}^{A_2^1} \bullet (\rho^{A_1^1} \otimes \mathbb{I}_{A_2^1}) \nonumber \\
&=\mathbb{I}_{A_1^1} \bullet \rho^{A_1^1} \label{eq:Idotrho}
\end{align}
where we have used the fact that for any trace-preserving channel $\mathbb{I}_B \bullet J_A^B = \mathbb{I}_A  $. Note that \eqref{eq:Idotrho} is independent of $J^{[1]}$ and hence does not depend on $x_1$. Using similar arguments for the other parties, $(J^{[1]} \otimes \ldots \otimes  J^{[N]}) \bullet  \eta $ must be a constant independent of the channels $J^{[1]}$ to $J^{[N]}$ (and hence independent of all of the inputs $x_1, \ldots ,x_N$). By appropriately normalising $\eta$ (which doesn't cause any physical effects in the two time state formalism) we can obtain 
\begin{equation} 
(J^{[1]} \otimes \ldots \otimes  J^{[N]}) \bullet  \eta =1
\end{equation} 
for all choices of channels $J^{[1]}$ to $J^{[N]}$. From the general rule for probabilities in the two time state formalism,
\begin{align}
p(a_1, \ldots ,a_N) &= \frac{(J^{[1]}_{a_1} \otimes \ldots \otimes  J^{[N]}_{a_N}) \bullet  \eta}{(J^{[1]} \otimes \ldots \otimes  J^{[N]}) \bullet  \eta } \nonumber \\
& = (J^{[1]}_{a_1} \otimes \ldots \otimes  J^{[N]}_{a_N}) \bullet  \eta
\end{align} 
which is the probability rule given in \eqref{eqn:defining}, and thus $\eta$ is a linear two-time state.

\section{Verification that $\eta$ is a Linear Two-Time State}
\label{append:etaistwotime}
Verifying that $\eta$ is linear two-time state amounts to showing that the relation 
\begin{equation} \label{eq:jkleta} 
(J \otimes  K \otimes  L )\bullet  \eta =1
\end{equation}
holds for all  trace-preserving channels $J$, $K$, and $L$. Given that $\eta$ contains a z-basis measurement in every link between parties, it is sufficient to consider only classical channels for the parties (see Appendix \ref{ap:classicalchannels}). For classical bits, there are a set of four channels that are extreme points of the convex set of classical channels- the identity, the flip or $X$ gate, and the throw away and replace with 0 and 1 channels. If \eqref{eq:jkleta}  holds for these it will hold for any convex combination of classical channels by linearity. Without loss of generality, consider Alice as the party of interest.

  Let $J$ be the identity channel firstly. Then consider 
\begin{equation}
(I_{A_i}^{A_o} \otimes  K_{B_i}^{B_o} \otimes  L_{C_i}^{C_o}) \bullet  \frac{1}{2}(M_{A_o} ^{B_i}\otimes M_{B_o} ^{C_i} \otimes M_{C_o} ^{A_i} + \bar{M}_{A_o} ^{B_i} \otimes \bar{M}_{B_o} ^{C_i} \otimes \bar{M}_{C_o} ^{A_i}).
\end{equation}
By using the identity channel on Alice to connect and contract the indices this is equal to  
\begin{equation}
(K_{B_i}^{B_o} \otimes  L_{C_i}^{C_o}) \bullet  \frac{1}{2}(M_{A_o} ^{B_i} \bullet M_{C_o} ^{A_o}  \otimes M_{B_o} ^{C_i} + \bar{M}_{A_o} ^{B_i} \bullet \bar{M}_{C_o} ^{A_o} \otimes \bar{M}_{B_o} ^{C_i} ).
\end{equation}
We have the relations operationally that $M_{A} ^{B} \bullet M_{B} ^{C}=M_{A} ^{C}$ and $\bar{M}_{A} ^{B} \bullet \bar{M}_{B} ^{C}=M_{A} ^{C}$ which gives  
\begin{align}
 &(K_{B_i}^{B_o} \otimes  L_{C_i}^{C_o}) \bullet  \frac{1}{2}(M_{C_o} ^{B_i}\otimes M_{B_o} ^{C_i} + M_{C_o} ^{B_i} \otimes \bar{M}_{B_o} ^{C_i} )\\
=&  (K_{B_i}^{B_o} \otimes  L_{C_i}^{C_o}) \bullet (M_{C_o} ^{B_i}\otimes \frac{1}{2}(M_{B_o} ^{C_i} +  \bar{M}_{B_o} ^{C_i})).
\end{align}
We note that $\frac{1}{2} \left(M_{B_o} ^{C_i} +  \bar{M}_{B_o} ^{C_i}\right)$ is equivalent to the operation throw away and replace with the maximally mixed state, i.e. has the operational form $\mathbb{I}_{B_o} \otimes \frac{1}{2}\mathbb{I}^{C_i}$. Since all the remaining channels are by definition trace preserving they satisfy $\mathbb{I}_B \bullet J_A^B = \mathbb{I}_A  $, so we can move the identity through the state giving 
\begin{align}
&\frac{1}{2} (K_{B_i}^{B_o} \otimes  L_{C_i}^{C_o}) \bullet (M_{C_o} ^{B_i}\otimes \mathbb{I}_{B_o} \otimes \mathbb{I} ^{C_i} ) \nonumber \\
=&\frac{1}{2} (\mathbb{I}_{B_i} \otimes  L_{C_i}^{C_o}) \bullet (M_{C_o} ^{B_i} \otimes \mathbb{I} ^{C_i} )\nonumber\\
=&\frac{1}{2} L_{C_i}^{C_o} \bullet (\mathbb{I}_{C_o} \otimes \mathbb{I} ^{C_i} )\nonumber\\
=&\frac{1}{2} \mathbb{I}_{C_i} \bullet \mathbb{I} ^{C_i}\nonumber \\
=&1.
\end{align}
We should also confirm this when Alice applies the flip channel;
\begin{equation}
X_{A}^{B}=|0 \rangle^\mathcal{B} \! \otimes\! \,_\mathcal{A}\langle 1|\otimes |1 \rangle ^{\mathcal{A}^\dagger}\! \otimes \!\,_{\mathcal{B}^\dagger}\langle 0| + |1 \rangle^\mathcal{B} \! \otimes\! \,_\mathcal{A}\langle 0|\otimes |0 \rangle ^{\mathcal{A}^\dagger}\! \otimes \!\,_{\mathcal{B}^\dagger}\langle 1|
\end{equation}
which gives
\begin{equation}
(X_{A_i}^{A_o} \otimes  K_{B_i}^{B_o} \otimes  L_{C_i}^{C_o}) \bullet  \frac{1}{2}(M_{A_o} ^{B_i}\otimes M_{B_o} ^{C_i} \otimes M_{C_o} ^{A_i} + \bar{M}_{A_o} ^{B_i} \otimes \bar{M}_{B_o} ^{C_i} \otimes \bar{M}_{C_o} ^{A_i}).
\end{equation}
Now we have the relations that
$X_{A}^{B} \bullet  M_{B}^{C}= \bar{M}_{A}^{C}$
and
$X_{A}^{B} \bullet \bar{M}_{B}^{C}=M_{A}^{C}$.
Then we get 
\begin{align}
& ( K_{B_i}^{B_o} \otimes  L_{C_i}^{C_o}) \bullet  \frac{1}{2}(M_{B_o} ^{C_i} \otimes \bar{M}_{A_i} ^{B_i}  \bullet M_{C_o} ^{A_i} + \bar{M}_{B_o} ^{C_i} \otimes M_{A_i} ^{B_i}  \bullet \bar{M}_{C_o} ^{A_i})\nonumber \\
& = ( K_{B_i}^{B_o} \otimes  L_{C_i}^{C_o}) \bullet  \frac{1}{2}( M_{B_o} ^{C_i} \otimes \bar{M}_{C_o} ^{B_i} +  \bar{M}_{B_o} ^{C_i} \otimes \bar{M}_{C_o} ^{B_i})\nonumber \\
& = ( K_{B_i}^{B_o} \otimes  L_{C_i}^{C_o}) \bullet  ( \frac{1}{2}(M_{B_o} ^{C_i} +\bar{M}_{B_o} ^{C_i}) \otimes \bar{M}_{C_o} ^{B_i} )\nonumber \\
& = \frac{1}{2}( K_{B_i}^{B_o} \otimes  L_{C_i}^{C_o}) \bullet  ( \mathbb{I}_{B_o} \otimes \mathbb{I}^{C_i}  \otimes \bar{M}_{C_o} ^{B_i} )\nonumber \\
&=1.
\end{align}
By playing the same trick with all other channel types we can recover similar results. For instance the throw away and replace with zero channel has the operational form
\begin{equation}
\mathbb{I}_{A_i} \otimes |0 \rangle ^{\mathcal{A}_o}\! \otimes\! \,_{\mathcal{A}^{\dagger}_o} \langle 0|.
\end{equation}
the method here can be visualised easily - if we have a circular structure of future preserving channels, we can always move identity `backwards' around the circle, until it `hits' the prepared state and gives 1. In particular 
\begin{align} 
& (\mathbb{I}_{A_i} \otimes |0 \rangle ^{\mathcal{A}_o}\! \otimes\! \,_{\mathcal{A}^{\dagger}_o} \langle 0|\otimes  K_{B_i}^{B_o} \otimes  L_{C_i}^{C_o}) \bullet  \frac{1}{2}(M_{A_o} ^{B_i}\otimes M_{B_o} ^{C_i} \otimes M_{C_o} ^{A_i} + \bar{M}_{A_o} ^{B_i} \otimes \bar{M}_{B_o} ^{C_i} \otimes \bar{M}_{C_o} ^{A_i})
\nonumber \\
&=( |0 \rangle ^{\mathcal{A}_o}\! \otimes\! \,_{\mathcal{A}^{\dagger}_o} \langle 0|\otimes  K_{B_i}^{B_o} \otimes  L_{C_i}^{C_o}) \bullet  \frac{1}{2}(M_{A_o} ^{B_i}\otimes M_{B_o} ^{C_i} \otimes \mathbb{I}_{C_o}  + \bar{M}_{A_o} ^{B_i} \otimes \bar{M}_{B_o} ^{C_i} \otimes \mathbb{I}_{C_o})
\nonumber \\
&=( |0 \rangle ^{\mathcal{A}_o}\! \otimes\! \,_{\mathcal{A}^{\dagger}_o} \langle 0|\otimes  K_{B_i}^{B_o} \otimes  \mathbb{I}_{C_i}) \bullet  \frac{1}{2}(M_{A_o} ^{B_i}\otimes M_{B_o} ^{C_i}  + \bar{M}_{A_o} ^{B_i} \otimes \bar{M}_{B_o} ^{C_i} )
\nonumber \\
&=( |0 \rangle ^{\mathcal{A}_o}\! \otimes\! \,_{\mathcal{A}^{\dagger}_o} \langle 0|\otimes  K_{B_i}^{B_o} ) \bullet  \frac{1}{2}(M_{A_o} ^{B_i}\otimes \mathbb{I}_{B_o}  + \bar{M}_{A_o} ^{B_i} \otimes \mathbb{I}_{B_o} )
\nonumber \\
&=( |0 \rangle ^{\mathcal{A}_o}\! \otimes\! \,_{\mathcal{A}^{\dagger}_o} \langle 0|\otimes  \mathbb{I}_{B_i} ) \bullet  \frac{1}{2}(M_{A_o} ^{B_i} + \bar{M}_{A_o} ^{B_i}  )
\nonumber \\
&=( |0 \rangle ^{\mathcal{A}_o}\! \otimes\! \,_{\mathcal{A}^{\dagger}_o} \langle 0| ) \bullet  \mathbb{I}_{A_o} )
\nonumber \\
&=1.
\end{align} 

Hence, the state is a linear two time state for all four extreme classical trace-preserving channels, and thus $\eta$ is a linear two time state.

\section{Quantum channel between measurements} \label{ap:classicalchannels} 
 We observe the fact here that any quantum channel that is sandwiched between two $z$-measurements is effectively classical, and superposition is destroyed. Consider a quantum channel with Kraus operators $A_k$, `sandwiched' between two z basis measurements;
\begin{align} 
M_z[\mathcal{E}[M_z(\rho)]]=  \sum_{i,j,k}|j \rangle \langle j |A_k|i \rangle \langle i |\rho |i \rangle \langle i | A_k^{\dagger}|j \rangle \langle j | 
=  \sum_{i,j}q_{j|i}|j \rangle \langle i |\rho |i \rangle \langle j |
\label{eqn:mapbetweenmeasure}
\end{align}
where $q_{j|i}$ is equal to
\begin{equation}
q_{j|i}=\sum_{k} \langle j |A_k|i \rangle \langle i | A_k^{\dagger}|j \rangle.
\end{equation}
$q_{j|i}$ is clearly positive and real, being the magnitude of a complex number. We can identify \ref{eqn:mapbetweenmeasure} with a classical stochastic map if the numbers $q_{j|i}$ can be identified with probabilities, which they can be since
\begin{align}
\sum_{i}q_{j|i}=\sum_{k,i}\langle j |A_k|i \rangle \langle i | A_k^{\dagger}|j \rangle=\sum_{k}\langle j |A_k A_k^{\dagger}|j \rangle=\langle j |I|j \rangle=1.
\end{align}
Thus, any quantum channel for qubits that gets sandwiched between two z-basis measurements can be simulated by a classical channel.

\end{document}